\documentclass[a4paper]{jpconf}
\usepackage{graphicx}
\def\be{\begin{equation}}
\def\ee{\end{equation}}
\def\te{\end{equation}}
\def\bea{\begin{eqnarray}}
\def\ba{\begin{eqnarray}}

\def\ta{\end{eqnarray}}
\def\tea{\end{eqnarray}}

\def\ben{\begin{enumerate}}
\def\een{\end{enumerate}}

\def\ha{{1\over 2}}

\def\n{\nu}

\def\bfr{{\bf r}}
\def\bfx{{\bf x}}

\def\bfL{{\bf L}}

\begin{document}
\title{Probing a gravitational cat state: Experimental Possibilities}

\author{M. Derakhshani}
\address{New York, New York.}
\ead{maaneliD@yahoo.com}
\author{C. Anastopoulos}
\address{Department of Physics, University of Patras, 26500 Patras, Greece.}
\ead{ anastop@physics.upatras.gr}
\author{B. L. Hu}
\address{Maryland Center for Fundamental Physics and Joint Quantum Institute,\\ University of
Maryland, College Park, Maryland 20742-4111 U.S.A.}
\ead{blhu@umd.edu}

\begin{abstract}
This is a progress report  on a preliminary feasibility study of experimental setups for preparing and probing a gravitational cat state \cite{ProbeCat2014}.
\end{abstract}

\section{Introduction}
As a follow-up to the theoretical studies of \cite{ProbeCat2014}, this short note opens the explorations for the best suited  schemes for the making and probing of a gravitational cat (g-cat) state,  using the currently available experimental proposals.  In the nature of a progress report, we aim here to share our thoughts for further discussions, leaving plenty of room for improvements and  broader collective wisdom.

\subsection{Gravitational cat states}


Consider the quantum description of a  stationary point mass $M$  localized around $\bf x = 0$ with spread $\sigma$, described by a Gaussian wave function with zero mean momentum.
 \begin{eqnarray}
\psi_0(\bfx) =  \frac{1}{(2 \pi \sigma^2)^{3/4}} {e^{ - \frac{{\bfx}^2}{4\sigma^2}} }. \label{Gausswfn}
\end{eqnarray}
The position ${\bfx}$ of the particle is a random variable described by the probability distribution $|\psi_0(\bfx)|^2$. According to Newton's law, a probability distribution for ${\bf x}$ defines a probability distribution for the Newtonian force acted on a particle of mass $m$ located at ${\bf R}$
\begin{eqnarray}
{\bf F} = - \frac{GMm}{|\bf R - \bf x|^3}(\bf R - \bf x). \label{newtforce}
\end{eqnarray}
 For $|{\bf R}| >> \sigma$ the fluctuations of the Newtonian force are negligible,  which leads one to view it as a deterministic variable.


 Now consider a cat state, i.e., a superposition of two Gaussians, each  located at $\pm \ha {\bf L}$ and  with zero mean momentum,
 \begin{eqnarray}
\psi(\bfx) = \frac{1}{\sqrt{2}} \frac{1}{(2 \pi \sigma^2)^{3/4}} \left[e^{ - \frac{(\bfx + \bfL/2)^2}{4\sigma^2}} +     e^{ - \frac{(\bfx - \bfL/2)^2}{4\sigma^2}}         \right]  \label{catwfn}
\end{eqnarray}
If ${\bf L}$ is of the order of magnitude of ${\bf R}$, the fluctuations of the Newtonian force (\ref{newtforce}) are non-negligible. Since the force is a function of ${\bf x}$, and ${\bf x}$ is described by an operator in quantum mechanics, the Newtonian force should also be described as an operator. But then, so would be the gravitational potential.  In this sense, the cat state for the point mass has generated a cat state for the gravitational field.


The model presented in \cite{ProbeCat2014} for a gravitational cat state involves a  quantum particle of mass $M$ confined in a symmetric  potential, as in Fig. 1.  The potential has two local minima located at $\bfr = \pm \frac{1}{2} \bfL$. We label the minima as $+$ and $-$.  (At a macroscopic level of observation,  the particle only lies in the $+$ region or in the  $-$region.)  With $|+ \rangle$ and $|-\rangle$ as the states localized around the minima $+$ and $-$ respectively,  the most general state is given by
\begin{eqnarray}
|\psi \rangle = c_+ |+\rangle + c_- |-\rangle.
\end{eqnarray}
  We assume a Hamiltonian $\hat{H} = \nu \hat{\sigma}_1$, where $\nu$ is a small, but non-vanishing, tunneling rate between the two minima.

\begin{figure}[h!]
\centering
\includegraphics[width=0.4\textwidth]{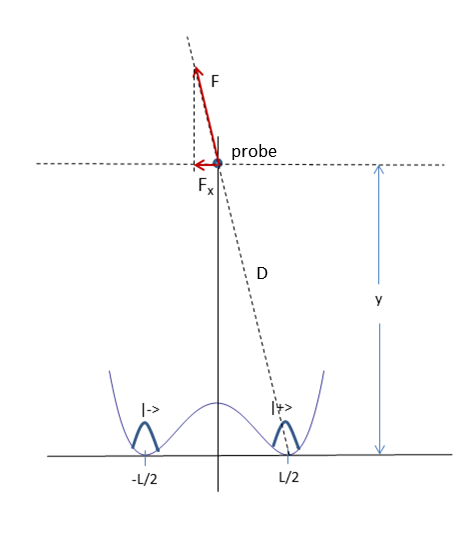}

\caption{Force on a probe  exerted by a massive particle in a gravitational cat state, $c_{+}|+>+c_{-}|->$.}
\
\end{figure}

Then we consider two ways of probing the gravitational field generated by the massive object in a g-cat state.

\subsection{A classical probe}
  We consider  a test  mass $m$ located near the confining potential, in a geometry described by Fig. 1.
Assuming that the  probe/detector is not allowed to move, the force $F$ in the horizontal direction takes only two values $f_0$ and $-f_0$, where
 \begin{eqnarray}
f_0= \frac{GMmL}{2D^3} \label{f0},
 \end{eqnarray}
 where $D = \sqrt{y^2+L^2/4}$ is the distance between the potential minimum and the location of  the probe; $y$ is shown in Fig. 1.

 We found  that for an initial  $|+\rangle$ state, the expectation value of $F$ and its two-time correlation function are given by
 \begin{eqnarray}
\langle F(t)\rangle = - f_0 e^{-\Gamma t} \label{corrcon1}
\\
\langle F(t') F(t)\rangle = f_0^2 e^{- \Gamma |t'-t|}. \label{corrcon2}
\end{eqnarray}
 The decay constant $\Gamma$ is defined as
 \begin{eqnarray}
  \Gamma = \frac{\nu^2 \tau}{2},
 \end{eqnarray}
where $\tau$ is the temporal resolution of the probe.

\subsection{A quantum probe} 
 The quantum probe invoked in \cite{ProbeCat2014} is a harmonic oscillator of mass $m$ and  frequency $\omega$ that is constrained to move along the horizontal axis of Fig. 1. If the amplitude of the oscillations is much smaller than $L$, the length scale of the cat state,   the force acted upon the oscillator along the $x$ direction  equals $\pm f_0$.
 Thus, the total Hamiltonian of the quantum massive object in a g-cat state interacting with the quantum oscillator probe is
\begin{eqnarray}
\hat{H} =   \n \hat{\sigma}_1 + \omega \hat{a}^{\dagger} \hat{a} + g \hat{\sigma}_3  (\hat{a} + \hat{a}^{\dagger}), \label{hjc}
\end{eqnarray}
where
\begin{eqnarray}
g = - \frac{f_0}{\sqrt{2m \omega}}.
\end{eqnarray}
We note that this  is the Hamiltonian of a single-mode Jaynes-Cummings model.

The oscillator can act as a probe of the gravitational cat only if
\begin{eqnarray}
 \left( \frac{f_0}{m}\right)^2 \frac{m}{\omega^3} >> 1. \label{dist}
\end{eqnarray}
 This corresponds to the ultra-strong coupling limit of the Jaynes-Cummings model, which is the physically relevant regime.
In this limit, the oscillator probe may undergo two types of quasi-classical oscillatory motion centered either at  $x_0 = \frac{f_0}{m \omega^2}$ or at  $x_0 = -\frac{f_0}{m\omega^2}$. The non-vanishing tunneling rate induces transitions between the two oscillatory motions and thus generates Rabi-type  oscillations with frequency $\nu$.

\section{Preparing a GravCat State}
In what follows, we consider the best proposals in preparing a g-cat state. In the next section, we discuss the experimental schemes best suited to the detection of a g-cat state.

In our present assessment, the most promising experimental proposal for the preparation of a g-cat state is  Romero-Isart et al.'s superconducting microsphere trapped in a harmonic potential created by a magnetic quadrupole field \cite{Romero-Isart2012}.
Having considered various sources of environmental decoherence in their set up, they suggest that this trapping method should make it possible to
isolate a lead (Pb) microsphere of
\begin{eqnarray}
\mbox{ mass} \hspace{0.2cm} M\sim10^{14} amu \hspace{0.2cm} \mbox{and radius} \hspace{0.2cm} R=2 \mu m \label{estimateRI}
\end{eqnarray}
to a degree sufficient enough to place the microsphere in a coherent superposition of two position eigenstates. The protocol
they propose for creating this microsphere cat state is parametric
coupling to a qubit state.

The estimate of the mass and radius of the microsphere by Romero-Isart et al. is obtained as follows.
The (superconducting) microsphere is trapped in
a 3-D harmonic oscillator potential of the form

\begin{equation}
V_{trap}=\frac{M}{2}\left[\omega_{t}^{2}\hat{x}^{2}+\omega_{\perp}^{2}(\hat{y}^{2}+\hat{z}^{2})\right],
\end{equation}
where the trapping frequency

\begin{equation}
\omega_{t}\simeq \frac{\left(\sqrt{\mu_{0}/\rho}\right)I}{l^{2}},
\end{equation}
and $\omega_{\perp}=\frac{\omega_{t}}{2}$, because the potential
is created by a quadrupole magnetic field that traps the microsphere
via the Meissner effect. The mass density $\rho$ is assumed to be a constant.
The parameter $l$ is the radius of and separation between the anti-Helmholtz
coils surrounding the microsphere as illustrated in Figure 2-a of \cite{Romero-Isart2012}, and $I$ is the current
through the coils.

 The microsphere can be trapped if the magnetic field at any point of the sphere is
smaller than the critical field, $B_{crit}$, in order to allow superconductivity.
This yields an upper bound on the radius of the sphere as

\begin{equation}
R<R_{max}\simeq \frac{B_{crit}}{\omega_{t}\sqrt{\mu_{0}\rho}}.
\end{equation}

The radius $R$ must also be much larger than the sphere's penetration length $\lambda$ and the coherence length $\xi$.
 The estimate (\ref{estimateRI}) is obtained from the following choice of parameters:
$\rho=11,360\frac{kg}{m^{3}}$, $\lambda=30.5nm$,
$\xi=96nm$ (at T = 0), $B_{crit}=0.08T$, $l=25\mu m$,
$I=10A$, and 
$\omega_{t}\simeq2\pi\times28kHz$. With these parameters, the maximum radius $R_{max}$ is about $3.7\mu m$.




With this  they show that parametric coupling to a qubit state puts the
microsphere in a spatial superposition   described by the wave function

\begin{equation}
|\Psi_{s}>=\frac{1}{\sqrt{2}}\left[\hat{T}(-2\chi)|\uparrow,0>+\hat{T}(2\chi)|\downarrow,0>\right],
\end{equation}
 where $\chi$ is a dimensionless parameter that characterizes the
parametric coupling, and $\hat{T}(...)$ is the usual translation
operator. The distance $L$ between the two superposed wave packets is $L = 4\chi x_{zp}$, where 
  $x_{zp}$ is the zero-point motion of the microsphere in the trap. From the values Romero-Isart et al. give for $\chi$ and $x_{zp}$, it can be readily confirmed that $L\sim 10^{-12} m$.

\section{Probing a GravCat state}

One of us (MD) has examined the leading state-of-the-art proposals in the past five years for ultrasensitive force measurement. A survey of which is contained in the Appendix.   Our analysis below will focus on  classical probes.

\subsection{Classical Probe}

For the role of the classical probe, the most promising experimental
proposal appears to be Reinhardt et al.'s "trampoline" resonator \cite{Sankey2015} made of Si$_{3}$N$_{4}$,
with effective mass $m = 4.0 ng$, width $100 \mu m$,
and projected force sensitivity of $\sim14 zN$ at cryogenic temperatures
($14 mK$).



While the resonator is a square-like membrane rather than a point particle, the latter assumed in the model of \cite{ProbeCat2014}, we can employ Eq. (\ref{f0}) for an order of magnitude estimate of the force. For a resonator of mass $m = 4.0 ng$, a microsphere of mass $M = 0.38 ng$, $L = 1pm$ and $D = 3\mu m$ ($1 \mu m$ larger than the radius of the microsphere of \cite{Romero-Isart2012}), we obtain
\begin{equation}
f_{0} =\frac{GmML}{2D^3}
 \sim 2\times10^{-30}N, \label{estimateF1}
\end{equation}
which is about ten orders of magnitude out of reach from the projected force sensitivity range of the resonator.



In order to examine possible ways to enhance the resonator--microsphere
gravitational interaction, we write the distance $D$ as $R + a$, where $R$ is the radius of the microsphere, which can be made variable, and $a$ is a fixed distance between the surface of the sphere and the resonator---we will consider $a$ to be of the order of one micrometer. Then
\begin{eqnarray}
f_0 \simeq (2) \frac{G\rho_m m L}{(1 + a/R)^3},
\end{eqnarray}
where $\rho_m = M/(\frac{4}{3} \pi R^3)$ is the density of the microsphere.

\begin{enumerate}
\item The most important parameter for increasing the gravitational interaction $f_0$ is the size $L$ of the cat state, since $f_0$ is directly proportional to $L$. Note, however, that in the scheme of \cite{Romero-Isart2012}, $L$ is indirectly constrained by other variables, including the radius $R$ of the microsphere.


    \item There is also a more modest increase of $f_0$ with the radius $R$ of the microsphere (and hence with the mass $M$ of the microsphere). However, the value of $R$ is constrained from the experimental set-up, because the magnetic field at any point on the sphere must be smaller than
the critical field, $B_{crit}$, so that the Meissner state is preserved. The corresponding   gradient $b_{max}=\frac{B_{crit}}{R}$ is proportional to the trapping frequency $\omega_t$. The latter must be at least of the order of tens of kHz, in order to allow cooling of the center of mass to the ground state \cite{Romero-Isart2012}. Taking these constraints into account, it can be readily confirmed that the absolute upper limit to $R_{max}$ for a Pb microsphere is about $8\mu m$.

\item Decreasing $a$ would slightly increase the force, but below a certain value, Casimir forces may become non-negligible. For example, Mohideen and Roy \cite{Mohideen} measured the Casimir force on the pN scale between a metallized sphere of radius $\sim 100 \mu m$ and a flat plate of diameter $1.25 cm$, with sphere-surface separations from 0.9 to 0.1 $\mu m$. Thus, even though the microsphere (made of Pb) and membrane (made of Si$_3$N$_{4}$) are much smaller in size than the sphere and plate used in \cite{Mohideen} ,  the microsphere-membrane Casimir force (to whatever extent present) may be many orders of magnitude greater than the microsphere-membrane gravitational force, when $D - R $ becomes smaller than  $1 \mu m$. If so, this would seem to put a practical lower bound on  $a \geq 1\mu m$.

\item Choosing a superconducting element with a much larger density may increase the force by a factor of about 2. A good choice is the (Type-I superconductor) Tantalum with density $16.7 \frac{g}{cm^{3}}$,  and a critical field slightly larger than  Pb (thus a slightly larger value of $R_{max}$).
\end{enumerate}


 Assuming that we can increase the size of the cat $L$ by one order of magnitude, and expecting that $R = 5 \mu m$ is feasible for a Tantalum microsphere we obtain
 \begin{eqnarray}
 f_0 = 0.6 \times 10^{-28} N,
 \end{eqnarray}
which is still about eight orders of magnitude from present level of detector sensitivities.

An obvious possibility would be to increase the mass of the probe $m$. However, this would mean increasing the area of the membrane and other factors would have to be taken into account. In particular, the   \textit{gravitational self-energy }of the probe may be  of the same order of magnitude  as the interaction energy between the g-cat and the probe.  The effects of gravitational self-energy in this setting is an issue worthy of more careful considerations.

\subsection{Quantum Probe}
A quantum probe of the gravitational cat state has to satisfy the constraint (\ref{dist}).
In considering possible experimental implementations of this
probe, one of us (MD) found that the most promising candidate seems to be the state-of-the-art
optomechanical harmonic oscillator described in \cite{Aspelmeyer2014}.
Such an oscillator has a mass $m =  100 ng$, and would experience a Newtonian
gravitational force of $ 10^{-21}$ Newtons from a
Pb microsphere cat state.
However, the dimensionless quantity of Eq. (\ref{dist}) is of the order of $10^{-53}$. This means that
  even with state-of-the-art optomechanical oscillators,
we are still a long way off from experimentally realizing the quantum
probe of a gravitational cat state.

\section{Conclusions}

Here we discuss the key findings of this report and related (theoretical) questions under study.

\subsection{Appraisal}

In this report we have examined the possibility of realizing and detecting a gravitational cat at the lab. From existing experimental proposals based on current and reachable technology we have  identified a set-up which can best create a superposition of macroscopically distinct states and can best measure the ultraweak gravitational force involved.  

In our preliminary assessment the measurement of a gravitational cat state is ten orders of magnitude from present capabilities, a difference that can perhaps be trimmed to eight orders of magnitude with relatively small improvement. More promising, perhaps, is an impending proposal by Romero-Isart and his collaborators \cite{Romero-IsartPrivate} to use free wave packet expansion in a "skatepark" potential to create coherent microsphere cat states (of the same mass) with an $L$ on the order of hundreds of nanonmeters. With this increase in $L$ the above force estimates would increase by five orders of magnitude or more, i.e., $\sim 10^{-25} N$ for the preliminary assessment and $\sim10^{-23} N$ with small improvement. Of course, these are optimistic estimates, as the combination of two set-ups, one for the creation of a cat state and one for the measurement of the force, will probably create unforeseen constraints on the main parameters of the experiment.  What we have not examined is how to reproduce the specific predictions in the gravitational cat models of Ref. \cite{ProbeCat2014}, which involve the crucial feature of tunneling between the distinct macroscopic configurations of the cat.

The preliminary estimates we presented above seem to suggest that  the  quantum effects of a matter source manifested through its gravitational field interactions could become measurable in the next (or next-next) generation of experiments.  Hence, it is worth exploring improved designs built upon our simple theoretical prototype. The aim is to  maximize the strength of the interaction between matter in the  cat state and the probe,  in order to  make  g-cat effects measurable.

\subsection{Open theoretical issues}

There are many additional theoretical issues related to both the gravitational cat state and the probes, e.g.,  1) how intact a gravcat state could remain, how long it could exist, in the presence of massive objects (such as the Earth); 2) why the gravitational force interaction with a classical probe should necessarily collapse the cat state wave function (in contrast to other massive bodies in nature, such as the Earth).

\subsection{Implications for alternative quantum theories}

In the theoretical model of the cat state used in \cite{ProbeCat2014}, i.e., standard GR+QFT, it was \emph{not} assumed that gravity is fundamentally quantum, but it was assumed that force measurements by a classical probe cause the gravitational field of the cat state to undergo quantum jumps (via quantum jumps of its mass density). Do alternative quantum theories of the objective collapse \cite{Bassi2013, Adler2013, Derak2014, TilloyDiosi2016}, hidden-variables \cite{Goldstein2013, Struyve2015}, and many-worlds \cite{Vaidman2014, PageGeilker1981} type agree with or contradict this assumption, when extended to the domain of semi-classical gravity? This requires a detailed discussion, to be given elsewhere.

\section*{Appendix: Experimental Schemes for Classical Probes of a GravCat State}\setcounter{section}{1}

This Appendix contains a summary of the search by one of us (MD) for the best experimental schemes which can  function as classical probes of gravitational cat states.  Amongst the handful of state-of-the-art proposals in the past five years for ultra-sensitive force measurements listed below he was able to  to identify only one ultra-sensitive force measurement scheme that
can play the role of a classical probe of sufficiently large mass
and sufficiently high force-measurement sensitivity which, in
combination with Romero-Isart et al.'s superconducting microspheres proposal  \cite{Romero-Isart2012}, could lead to an experimental scheme for measuring g-cat effects..

The six candidate proposals were:

\begin{enumerate}

\item Schreppler et al.'s scheme involving an ultra-cold atom cloud in
a high finesse cavity \cite{Schreppler2014} which, to date, produced
the smallest externally applied force measured of $42 yN$.

\item Moser et al.'s scheme involving carbon nanotube mechanical resonators
with quality factors greater than a million \cite{Moser2013, Moser2014},
which yields force measurements on the zN scale.

\item Tao et al.'s scheme using single-crystal diamond nanomechanical
resonators with quality factors exceeding one million \cite{Tao2014},
which produces force sensitivities of a few hundred zN.

\item Ranjit et al.'s scheme involving laser-cooled silica microspheres
as force sensors in a dual beam optical dipole trap in high vacuum
\cite{Ranjit2015}, which yields force measurement sensitivity
at the aN  scale.

\item Kleckner et al.'s \cite{Kleckner2011} and Reinhardt et al.'s \cite{Sankey2015}
schemes involving the use of optomechanical trampoline resonators,
which yield projected maximum force sensitivities on the aN and zN
scales, respectively.

\item Wagner et al.'s use of state-of-the-art torsion balance pendulums
to test for  violations of the weak equivalence principle with a precision of one part in $10^{13}$
\cite{Wagner2012}.

\end{enumerate}

Of all these, proposal 5 seems the most promising for our purpose. The justification for this is the following.

\textsc{Proposal 1} yields the greatest force sensitivity, but the atom cloud
used has a miniscule mass of only $1.8\times10^{-22} kg$. Using the center of mass of this atom cloud in place of the
resonator in Eq. (\ref{f0}) (and keeping the other parameters the same as those used in Eq. (17) of subsection 3.1), one obtains a gravitational force of  $\sim10^{-40} N$, which is seventeen orders of magnitude smaller than the force sensitivity of Schreppler
et al.'s scheme. In addition, the scheme requires that the measured force be
an externally applied force that oscillates at the natural frequency
of the center of mass motion of the atom cloud in the cavity ($\sim12
kHz$). It is unclear how this could be done with the gravitational
force from a cat state, even if it were somehow possible to fashion
an atom cloud with a center of mass of $\sim90 mg$.

\textsc{Proposal 2} has a similar obstacle in that the mass of these carbon
nanotube mechanical resonators is only $\sim10^{-20} kg$. Using this mass
  in Eq. (\ref{f0}), under the point mass approximation, yields a gravitational force of $\sim10^{-39} N$, or
eighteen orders of magnitude smaller than the force sensitivity of
the Moser et al. scheme.

\textsc{Proposal 3} used several different kinds of single-crystal diamond
nanomechanical resonators, the largest of which has a mass of $\sim10^{-12} kg$.
When the largest resonator is used in place of the resonator in Eq. (\ref{f0}), under the point mass approximation, it yields a gravitational force of $\sim10^{-31} N$. Since this resonator has a force sensitivity of only $540 zN$, the
expected gravitational force is around twelve orders of magnitude smaller.

\textsc{Proposal 4} uses silica microspheres of 3 micron diameter, with an
estimated mass of at most $\sim10^{-12} kg$. Thus these
microspheres yield a gravitational force around thirteen orders of magnitude
smaller than the force sensitivity of Ranjit et al.'s scheme.

\textsc{Proposal 6} is, of course, a different force measurement scheme
than all the others in that the goal of a torsion balance is to detect
a difference in the directions of the external force vectors applied
to the test bodies, rather than the absolute magnitudes of the forces.
The scheme described in Wagner et al. uses eight test bodies of masses
$5 g$ each and has a force sensitivity of one part in $10^{13}$ (for
the E{\"o}tv{\"o}s parameter), with a differential acceleration resolution of $\sim 10^{-15} m/s^{2}$. To compute the hypothetical (horizontal) gravitational force/acceleration
between one of these test bodies and the Romero-Isart et al. microsphere
cat state (i.e., using the latter as the attractor for the torsion balance), we need to know the size of one of these test bodies so that we can calculate the $D$ variable in Eq. (5). Although Wagner et al. do not specify the size or geometry of the test bodies, we can approximate the bodies as spherical mass distributions and deduce their radii from knowing what elements (hence mass densities) compose them. They state that four of the test bodies are made of element Be, the other four made of either Ti or Al. Let us consider a test body made of the element with the largest mass density (since, for fixed surface separation, $a$, this will give the smallest $R$ value hence the largest force). Among the three, Ti has the largest mass density with $\rho_{Ti} = 4.5g/cm^{3}$. The corresponding radius is then $R_{Ti} = 0.24 cm$. Assuming $a = 1 \mu m$, then $D = R_{Ti} + R_{sphere} + a \sim 10^{-3} m$. Using this $D$ value in Eq. (5), we obtain a force $\sim 10^{-29} N$, or a horizontal acceleration of $\sim10^{-28}m/s^{2}$ for one test body. This is around thirteen orders of magnitude smaller than the maximum sensitivity of the torsion balance. 
Of course, for such macroscopic test bodies, it seems implausible that experimentalists could arrange $a = 1 \mu m$; the experimental setup in Romero-Isart et al. \cite{Romero-Isart2012} (Figure 1-a therein) involves surrounding the microsphere by an anti-Helmholtz
coil configuration only 25 microns in width. Much more experimentally
feasible, it seems, is a surface separation on the order of a centimeter
or possibly a millimeter, either of which only further decrease the magnitude of the force/acceleration.

Why then does proposal 5 seem the most promising for our purpose? First, the largest trampoline
resonator used in the Kleckner et al. proposal has a mass of $110
ng$, diameter of 80 microns (see Figure 2 in \cite{Kleckner2011}),
and a projected force sensitivity on the aN scale at cryogenic temperatures.
To be more precise about this last feature, Kleckner et al. write:

\begin{quote}

Trampoline resonators are also suitable for use as ultra-high resolution
force sensors. Assuming the quality factor increase is also seen for
the lowest frequency devices {[}i.e., the resonator with \emph{m}
= 110 ng{]}, it should be possible to obtain a thermal force noise
in the aN/Hz regime at demonstrated [cryogenic] temperatures.
This is comparable to or better than the single crystal Si resonators
currently used in magnetic resonance force microscopy (MRFM) experiments... Furthermore, the rear side optical access can be used
to provide extremely precise position sensitivity while leaving the
front side free for surface modifications required for use as sensors.
(Last page)

\end{quote}
Now, for the gravitational force between the $110 ng$ trampoline resonator (approximated as a point mass)
and the microsphere, Eq. (5) gives $\sim5\times10^{-29} N$. By comparison to the previous proposals, this is 'only' ten to eleven orders of magnitude away from the projected force sensitivity range of their resonator. Second, the largest trampoline resonator used in the Reinhardt et al. proposal \cite{Sankey2015} has a mass of $4.0 ng$, width of 100 microns
(Figure 1 therein), and a projected force sensitivity of $\sim14 zN$ at cryogenic temperatures
($14 mK$). For the gravitational force between the $4.0 ng$ resonator and
the microsphere, Eq. (5) gives $\sim2\times10^{-25} N$, or about ten orders of
magnitude away from the projected force sensitivity range of the resonator.

Because the Reinhardt et al. resonator yields a force closest to its projected maximum force sensitivity (for the assumed parameters), and because Reinhardt et al. are more specific than Kleckner et al. in regards to the magnitude and conditions of maximum force sensitivity, their resonator was chosen as the most
promising force probe to combine with Romero-Isart et al.'s proposal. \\

\noindent {\bf Acknowledgment}  BLH  and MD wish to thank Gerhard Gr\"ossing and Jan  Walleczek for their regal Viennese receptions at the EmQM (Emergent Quantum Mechanics) 2015 meeting. BLH also wishes to thank Philip Stamp and organizers of the second PITP- Galiano Island meeting (Probing the mystery: Theory \& Experiment in Qunatum Gravity) in August 2015 for their eco-immersed Pacific-Northwest  hospitality.  The Galiano discussions amongst leading experimentalists working at the interface of quantum and gravitation physics  provided a good stimulus for us to explore the experimental possibilities  in the  realization of the theoretical scheme proposed  in \cite{ProbeCat2014}.  We are particularly grateful to Oriol Romero-Isart for the many valuable correspondences about his proposed superconducting sphere experiment.


\end{document}